\documentclass[12pt]{article}

\usepackage{amssymb,amsmath}
\usepackage{latexsym}
\usepackage{graphicx}
\usepackage{amsfonts}
\usepackage{amssymb}
\usepackage{epsfig}
\usepackage{color}
\usepackage{psfrag}

\newtheorem{rema}{Remark}[section]

\newtheorem{propo}[rema]{Proposition}

\newcommand{\bc}{\begin{center}}
\newcommand{\ec}{\end{center}}
\def\ba#1{\begin{array}{#1}\displaystyle}
\newcommand{\ea}{\end{array}}
\newcommand{\z}{\\ \displaystyle}
\newcommand{\beq}{\begin{equation}}
\newcommand{\eeq}{\end{equation}}
\newcommand{\beqa}{\begin{eqnarray}}
\newcommand{\eeqa}{\end{eqnarray}}
\newcommand{\no}{\nonumber}
\newcommand{\n}{\nonumber\\}
\newcommand{\bi}{\begin{itemize}}
\newcommand{\ei}{\end{itemize}}
\def\mato#1{\left(\ba{#1}} 
\def\matf{\ea\right)}
\def\cqfd{ {\hfill{$\Box$}} }

\def\lt#1{\left#1}
\def\rt#1{\right#1}
\def\t#1{\tilde{#1}}
\def\h#1{\hat{#1}}
\def\b#1{\bar{#1}}
\def\frc#1#2{\frac{#1}{#2}}

\newcommand{\p}{\partial}

\newcommand{\vac}{{\rm vac}}
\newcommand{\bra}{\langle}
\newcommand{\ket}{\rangle}

\newcommand{\R}{{\mathbb{R}}}
\newcommand{\C}{{\mathbb{C}}}
\newcommand{\Tr}{{\rm Tr}}

\newcommand{\Or}{{\cal O}}

\newcommand{\ep}{\epsilon}

\newcommand{\cur}{{\cal J}}

\begin{document}

\pagestyle{empty}
\bc {\Large Full Counting Statistics in the Resonant-Level Model}
\vskip 1.0 truecm

Denis Bernard${}^{\clubsuit}$\footnote{Member of C.N.R.S.; \texttt{denis.bernard@ens.fr}}
and
Benjamin Doyon${}^{\spadesuit}$\footnote{\texttt{benjamin.doyon@kcl.ac.uk}}
\ec

\vskip 0.5 truecm

\noindent
${}^{\clubsuit}$ Laboratoire de Physique Th\'eorique de l'Ecole Normale Sup\'erieure, CNRS/ENS, Ecole Normale Sup\'erieure, 24 rue Lhomond, 75005 Paris, France.\\
${}^{\spadesuit}$ Department of Mathematics, King's College London, Strand, London WC2R 2LS, United Kingdom.
\vskip 2 truecm

We derive the large deviation function, which provides the large-time full counting statistics for the charge transfer, in the non-equilibrium steady state of the resonant-level model. The general form of this function in free fermion models, in terms of transmission coefficients, was proposed by Levitov and Lesovik in 1993 using a particular measurement set-up involving an interacting spin. It was later suggested to hold as well for a proper quantum mechanical measurement of the transferred charge. We give a precise proof of both statements in the resonant-level model. We first give a full description of the model and its steady state. That is, we explain how the decoupled system prepared with a charge differential evolves, with the impurity coupling, towards the Hershfield non-equilibrium density matrix, in the sense of averages of finitely-supported operators. We describe how this holds both for the usual resonant-level model with a point-like impurity, and for a regularised model with an impurity spread on a finite region, shedding light on subtleties associated to the point-like impurity. We then prove Levitov-Lesovik formula by recasting the problem into calculating averages of finitely-supported operators.

\newpage

\pagestyle{plain}

\section{Introduction}
Full counting statistics in electronic transport refers to the probability distribution function (p.d.f.)~of charges transferred through quantum wires. In the simplest setting, one may have in mind (or for real) a quantum dot coupled to (at least) two electronic reservoirs which are maintained at different potentials (and different temperatures) so that an electronic current flows through the dot from one reservoir to the other. The system is by construction out of equilibrium. Since quantum effects dominate, one has to describe the reservoirs -- and, of course, the dot -- quantum mechanically.

Let $q$ be the charge transferred during time $t$. Its p.d.f.~is fully characterised by the generating function $P(\lambda,t)$,
\[ P(\lambda,t) = \sum_q e^{i\lambda q}\, \Pi_q \]
with $\Pi_q$ the probability of observing a transferred charge $q$. A formula for $P(\lambda,t)$ at large time has been given by Levitov and Lesovik \cite{LL1}. However, one has to specify how the transferred charges are actually measured.

A first way to define the transferred charge during a time duration $t$ consists in first measuring at time $0$ the charge, say, in one of the reservoirs, letting the system evolve, and then measuring again the charge at a later time $t$. Let $Q$ be the charge operator and $P_q$ the projector onto the $Q$-eigenspace with eigenvalue $q$. If $\rho_0$ is the normalised system density matrix at time $0$, basic principles of quantum mechanics tell us that the generating function for the transferred charges is:
\beqa\label{P}
	P(\lambda,t) &=& \sum_{q,q_0}
	e^{i\lambda q}\  \Tr\lt(P_{q_0+q} U_{0;t} P_{q_0}\, \rho_0\, P_{q_0} U_{0;t}^\dag P_{q_0+q}\rt)
	\eeqa
where $U_{0;t}$ is the system evolution operator from time $0$ to time $t$. The sum over $q_0$ corresponds to the sum over the values of the charge measured at time $0$, while that over $q+q_0$ corresponds to the charge measured at time $t$. Note that this formula involves two projectors $P_{q_0}$ and $P_{q+q_0}$.

Definition (\ref{P}) is actually not the way Levitov-Lesovik \cite{LL1} chose to specify the charge transferred. Instead they define the measured transferred charge by modeling a measurement device made of a quantum spin coupled to the electronic current flowing through the quantum wire. A precise description of the measurement system is not necessary for our discussion but may be found in \cite{Lev}. We only recall that it leads to another definition for the generating function:
\beqa\label{P2}
	\widehat P(\lambda,t) &=& \sum_{q,q_0}
	e^{i\lambda q }\  \Tr\lt( P_{q_0+q} U_{0;t} P_{q_0}\,  \rho_0\   U_{0;t}^\dag \rt)
\eeqa
with, again, $\rho_0$ the system density matrix and $U_{0;t}$ is the evolution operator. This formula also involves the two projectors $P_{q_0}$ and $P_{q+q_0}$ but it differs form the previous ones by the fact that there is only one insertion of the projector $P_{q_0}$ and not two as in Eq.~(\ref{P}). The two formulae would coincide if the density matrix happens to commute with the charge operator, and thus with the projectors $P_q$. But this is generically not the case, and certainly not if the density matrix represents a steady state where the charge $Q$ is flowing. None of these formula correspond to $\log \Tr(e^{i\lambda(Q_t-Q)}\, \rho)$  which would be the analogue of the classical formula. 

The universal character of the p.d.f.~encoded in (\ref{P}) only emerges at large time duration $t$ and in a stationary regime, that is only when the system ``reservoirs plus dot" is in a stationary state described by a (non-equilibrium) density matrix $\rho_{\rm stat}$. Let $P_{\rm stat}(\lambda,t)$ be the corresponding generating function.
We are interested in evaluating the large deviation function $F(\lambda)$ defined by
\beq\label{F}
	F(\lambda) = \lim_{t\to\infty} t^{-1} \log P_{\rm {stat}}(\lambda,t).
\eeq

The Levitov-Lesovik formula \cite{LL1} is an expression for $F(\lambda)$ associated to the generating function (\ref{P2}), in the case where the electrons propagate in the quantum dot through one channel only. It reads:
\beq \label{LLformula}
F(\lambda) = \int \frac{d\omega}{2\pi} \log\Big( 1 + T(\omega)\Big[ n_1(\omega)(n_2(\omega)-1)(1-e^{i\lambda})+n_2(\omega)(n_1(\omega)-1)(1-e^{-i\lambda})\Big]\Big) 
\eeq
with $T(\omega)$ the transmission coefficient at energy $\omega$ and
\beq\label{occupation}
	n_j(\omega)=\frc1{e^{(\omega-\mu_j)/T_j}+1}
\eeq
the Fermi occupation numbers of the two leads. When the two reservoirs are at identical temperature, $T_1=T_2=T$, the Levitov-Lesovik large deviation function satisfies the fluctuation relation \cite{fluctu,Espo}: $F(\lambda)=F(-\lambda-iV/T)$ with $V=\mu_1-\mu_2$ the potential difference. 

Of course, there is already quite a number of papers dealing with the Levitov-Lesovik formula. The original paper \cite{LL1} as well as the very nice review by Levitov \cite{Lev} deal with the definition (\ref{P2}) of the transferred charge motivated by a model for the quantum measure apparatus. Recall that definition (\ref{P2}) does not coincide with the von Neumann prescription for the measure of the quantum charge at different time intervals. Levitov-Lesovik \cite{LL1,Lev} derivation is based on a scattering approach. References \cite{Klich,Schonh} also deal with definition (\ref{P2}) of the transferred charges either using a microscopic formulation \cite{Schonh} or a scattering approach combined with free fermion technology \cite{Klich}. References \cite{Avron,Espo} start from the von Neumann definition of the measured transferred charge but Ref.~\cite{Avron} assumes that the charge operator commutes with the density matrix and Ref.~\cite{Espo} uses a quantum master equation approximation. Up to our knowledge, the approach to the stationary regime in relation with the full counting statistics does not seem to have been analysed in the previous literature, except partially in Ref.~\cite{BenNatan}.

The aim of this paper is to provide a complete derivation of the Levitov-Lesovik formula in the framework of the resonant-level model (whose definition is given below), one of the simplest models for a quantum dot. Although the formula is already known for some time, the derivation in the framework of the resonant-level model contains a few subtleties which have never been thoroughly sorted out. We introduce a regularisation of the model in which the impurity is spread over a small distance. First we prove that the system reaches a stationary state, which requires taking large size and large initial preparation time limits in the appropriate order. This limit actually exists only for local operators (as was suggested in the context of the Kondo model \cite{BenNatan}), that is, operators localised around the quantum dot. The charge operator is not a local operator, since it measures the charge in a reservoir, so that it does not admit large size stationary limit. We however show that the operator involved in the transferred charge p.d.f.~is nevertheless a local operator and that it admits a stationary limit. We then use free fermion techniques to evaluate the large deviation function. We also show that the two ways of defining the measured transferred charge, (\ref{P}) and (\ref{P2}), lead at large time and in the stationary regime to identical large deviation functions.

Our approach is strongly based on local-operator methods, and in that differs significantly from previous approaches to derive the Levitov-Lesovik formula. The idea of local operators is crucial both for the explicit calculations, and for showing the existence of the stationary limit. On the other hand, techniques using scattering states may be more amenable to interacting, integrable models, for instance. We did not make any attempt to connect our derivation to the scattering-state approach, and hope to come back to this in the future.

The paper is organised as follows. In Section \ref{sectmodel}, we define the resonant-level model taking care of the subtleties introduced by an {\em a priori} point-like impurity, we discuss how to describe the stationary state using density matrices of the Hershfield type, and we show that the stationary state is reached in the appropriate large-time limit. In Section \ref{sectcharges}, we prove the Levitov-Lesovik formula for the resonant-level model, using local-operator and free-fermion techniques.

\noindent {\bf Acknowledgment}\\
BD would like to acknowledge Durham University, where part of this work was done.

\section{Non-equilibrium stationary state in the resonant-level model} \label{sectmodel}

\subsection{General discussion: model and the stationary state}

The resonant-level model describes a quantum dot coupled to two electronic reservoirs. Each of the reservoirs is modeled by massless fermions propagating on a line segment $[0,L]$; after unfolding, this becomes chiral fermions on the interval $[-L,L]$ with periodic boundary conditions. The dot is modeled as simply as possible by a unique level which can be either empty or occupied. The coupling is such that electrons may hop between the dot and the reservoirs with identical amplitudes for hoping in either of the two reservoirs. The Hamiltonian we consider is $H=H_1+H_2+H_d+H_\text{int}$ with $H_j$ the free fermion Hamiltonians describing the reservoir dynamics, $H_d$ the dot Hamiltonian, and $H_\text{int}$ the dot-reservoir coupling. Explicitly \cite{RLM}:
\beq\label{H}
	H = -i v_F\int_{-L}^L dx(\psi_1^\dag \p_x \psi_1 + \psi_2^\dag\p_x \psi_2) +
	\frc{\tau}{\sqrt{2}} ( (\psi^\dag_1(0)+\psi^\dag_2(0)) d + d^\dag (\psi_1(0)+\psi_2(0)))
	+ \ep\, d^\dag d.
\eeq
where $d^{\dag}$ is the fermion creation operator on the dot, $\{d^{\dag},d\}=1$, $\epsilon$ the dot energy, $\psi_j(x)$, $j=1,2$, are the fermionic operators of the reservoirs, $\{\psi_j^{\dag}(x),\psi_k(y)\}=\delta_{j;k}\delta(x-y)$, with periodic boundary condition $\psi_j(L)=\psi_j(-L)$, $v_F$ is the Fermi velocity and $\tau$ the hopping amplitude.

The system is placed out-of-equilibrium by preparing, say at early time $t_0<0$, the two reservoirs at different temperatures, $T_1$ and $T_2$, and different chemical potentials $\mu_1$ and $\mu_2$.  The initial system density matrix is $\rho_d\otimes\rho_{\rm th}$ with $\rho_{\rm th}$  the reservoir thermal density matrix, with temperature $T_j$ and chemical potential $\mu_j$, and $\rho_d$ the dot density matrix. Explicitly, $\rho_{\rm th}=\rho_1\otimes\rho_2$ with $\rho_j=Z_j^{-1}e^{-\beta_j(H_j-\mu_jN_j)}$, $\Tr(\rho_j)=1$, where $N_j=\int dx\, \psi_j^\dag(x)\psi_j(x)$ and $H_j=-i v_F\int_{-L}^L dx\,\psi_j^\dag \p_x \psi_j$ (here $\beta_j=1/T_j$). Upon shifting energies, we may take $\beta_1\mu_1 +\beta_2\mu_2 = 0$ without loss of generality; we will assume that this relation holds in the following.

\begin{figure}
\bc
\includegraphics[width=14cm,height=2cm]{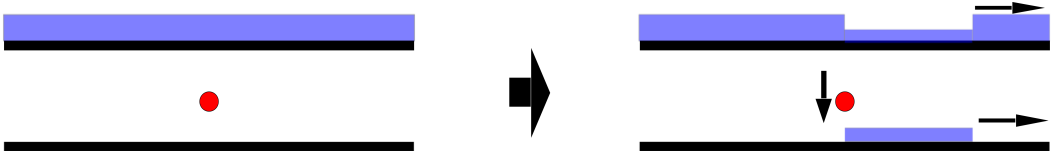}
\ec
\caption{A (cartoon) picture of the two (unfolded) electron baths and the impurity, and of the evolution of the system. The figure on the left represents the initial density matrix: the upper electron bath is filled and the lower one is empty, due to the initial potential difference. The large arrow indicates time evolution. The figure on the right represents the density matrix after a time $|t_0|$, evolved without a potential difference, along with the instantaneous motion of the electrons. They fall from the filled electron bath to the empty one in order to establish their zero-potential configuration, and the region of fallen electrons expands with velocity $v_F$. The steady-state limit is the one where the system is much larger than the region of fallen electrons, and where this region is much larger than the quantum correlation region around the impurity.}
\label{figlimit}
\end{figure}

Charge and heat currents are going to flow between the reservoirs through the dot. They roughly propagate at velocity $v_F$, so that a stationary regime is reached on a domain of scale $\ell_{\rm stat}$ (this is a quantum correlation length, here proportional to $\tau^2$) around the dot for large enough early time $|t_0|$ and large $L$  with 
\[ L\gg v_F|t_0| \gg \ell_{stat}. \]
The reservoir size has to be large enough for the charge and heat flows not to bounce back on the reservoir boundaries (see Fig. \ref{figlimit}).

Physics in the stationary domain is described by a stationary density matrix $\rho_{\rm stat}$ of the coupled system which, according to the previous discussion, is given by infinite $L$ and $t_0$ limits in the following order:
\beq\label{deflimdens}
	\rho_{\rm stat}= \lim_{t_0\to -\infty} \lim_{L\to\infty} U_{0;t_0}\, (\,\rho_d\otimes\rho_{\rm th}\,)\, U_{0;t_0}^{\dag}
\eeq
with $U_{t;t_0}=\exp(-i(t-t_0)H)$ the time-evolution operator from $t_0$ to $t$. This limit exists (in the weak sense) as long as the density matrix is evaluated against operators localised in the stationary domain $|x|\leq \ell_{stat} \ll v_F|t_0|$.

The stationary state density matrix can be described using Hershfield's density matrix \cite{Hersh} (the following is a slight generalisation, with $\beta_1\neq\beta_2$ in general):
\beq\label{rhostat}
	\rho_{\rm stat} = e^{-\beta H -WX + VY}
\eeq
where $\beta = (\beta_1+\beta_2)/2$, $V = \beta_1\mu_1-\beta_2\mu_2$, $W =\beta_1-\beta_2$, and the two operators $Y$ and $X$ are (generalisation of) Hershfield's operators describing the steady state. There are various ways one can formally define these operators. We will adopt that first used in \cite{Ben?}. It simply says that $Y\sim (N_1-N_2)/2$ and $X\sim (H_1-H_2)/2$ for $x<0$ (in general, for $x$ small enough), and that both $Y$ and $X$ are conserved by the dynamics. That is, we have $[Y,H]=[X,H]=0$ and:
\begin{eqnarray}\label{defYX}
	~\big[Y,\psi_1(x)\big] = -\frac{1}{2} \psi_1(x),& \big[X,\psi_1(x)\big] = \frc12 iv_F \p_x \psi_1(x),&\quad \text{for}\ x<0;\\
	~\big[Y,\psi_2(x)\big] = \frc12 \psi_2(x),& \big[X,\psi_2(x)\big] = -\frc12 iv_F \p_x \psi_2(x),&\quad \text{for}\ x<0.\nonumber
\end{eqnarray}
These specifications of the non-equilibrium density matrix apply to models defined as in Eq.~(\ref{H}) with a localised impurity. We shall introduce a regularised version of the model in which the impurity is spread over a small distance. The above relations (\ref{defYX}) will then be valid only for $x$ negative enough to be away from the impurity support, instead of all $x<0$. Note that $\rho_{\rm stat}$ describes a current flowing from lead 1 to lead 2. Physically, the conditions indicate that $\rho_{\rm stat}$ is stationary and that $\rho_{\rm stat}$ reproduces the initial density matrix at positions where the electrons have not yet interacted with the impurity, i.e.~before they flow from one lead to another. We will see below that there is a unique solution (up to the addition of terms proportional to the identity operator) to these conditions. Although the density matrix has a form that resembles that of an equilibrium density matrix, as was argued in \cite{Ben?,Hersh} it is the fact that the operators $Y$ and $X$ are non-local which makes it a non-equilibrium density matrix.

In the following, we set $v_F=1$. In Subsection \ref{ssect_sol}, we recall and develop further the solution to the resonant-level model, both by using a ``resolution'' of the impurity, and by using an explicit spreading of the impurity, in order to take care of the ambiguities produced by the impurity-interaction terms in the Hamiltonian (\ref{H}). In Subsection \ref{ssect_limit}, we prove that (\ref{defYX}) has a unique solution, and we prove (\ref{deflimdens}).

\subsection{The operator solution to the resonant-level model} \label{ssect_sol}

Let us now recall the solution of the resonant-level model (the diagonalisation of the Hamiltonian). We use the even and odd combinations
\[
	\psi_e(x) = \frc1{\sqrt2} (\psi_1(x) + \psi_2(x)),\quad \psi_o(x) = \frc1{\sqrt2} (\psi_1(x) - \psi_2(x))
\]
in terms of which the Hamiltonian decouples. As usual, the approach is to use a parametrisation of the operators $\psi_e(x)$, $\psi_o(x)$ and $d$ in terms of mode operators, such that the canonical anti-commutation relations hold and such that the time evolution of the mode operators is simple. We will look at various ways of regularising the impurity term present in the Hamiltonian (hence we will have various time evolutions).

\subsubsection{Partial regularisation via a resolution of the impurity}

The solution to the odd part is trivial as it does not involve the impurity (see (\ref{eqos}) below), hence let us discuss the even part. Since the Hamiltonian is quadratic in fermion operators, the full solution can be straightforwardly obtained from the one-particle solution. The one-particle (even) Hilbert space is the direct sum $L^2(\mathbb{S}) \oplus \C$: a square-integrable function $g(x)$ on the interval $[-L,L]$ with $g(-L)=g(L)$ representing the wave function of the even electron, and a complex number $h$ representing the amplitude for the presence of the electron on the impurity. There are two equations resulting from the eigenvector equation for $H$ with eigenvalue $p$:
\beqa
	-i\p_x g(x) + \tau h\, \delta(x) &=& p\, g(x) \n
	\tau g(0) + \ep\, h &=& p\, h.\label{schro}
\eeqa

Any solution to (\ref{schro}) is such that the jump is $g(0^+)-g(0^-) = -i\tau\, h$. Yet, clearly, the second equation necessitates the knowledge of $g(0)$, and this is also true for the evaluation of certain averages (e.g. the average of the momentum operator $-i\p_x$). A way of solving this problem is to use a resolution of the impurity point. We will replace $\psi_e(0)$ in the Hamiltonian (\ref{H}) by the symmetric sum $\psi_e(0)\equiv (\psi_e(0^+) + \psi_e(0^-))/2$, with the understanding that $0^\pm$ are limits from the right/left towards 0. The delta-function potential becomes likewise a symmetric sum of delta functions, which, under integrations with functions in general discontinuous at $x=0$, takes their symmetric sums. Also, the second equation becomes well-defined, since $g(0)$ is replaced by the symmetric sum of $g(0^+)$ and $g(0^-)$. This resolution is equivalent to defining $g(0)$ as $(g(0^+)+g(0^-))/2$ for averages of operators with up to one derivative, and exactly reproduces the limit where a spreading of the impurity tends to the delta-function (see the discussion below of the impurity spreading). We note that it is inconsistent to choose, as a resolution of $\psi_e(0)$, any non-symmetric sum of $\psi_e(0^+)$ and $\psi_e(0^-)$, as noted in \cite{Ben}; below we indeed show the universality of the chosen resolution: its independence from the way the impurity was spread. This resolution is not enough for an unambiguous definition of all averages, but it will be sufficient for our purposes.

The Hilbert space is obtained as a completion of the space of solutions to the resolved equations (\ref{schro}). A solution is obtained by an appropriate Fourier-like transform. We show in Appendix \ref{apphilbert} that the map $(a_p:p) \mapsto g\oplus h$ given by
\beq\label{basis}
	g(x) = \sum_p e_p(x)\, a_p,\quad
	-i\tau\, h = \sum_p w_p\, a_p,
\eeq
with
\beq\label{defep}
	e_p(x) := e^{ipx}  \times  \lt\{\ba{ll} 1& {\rm for}\ x<0 \z  v_p  & {\rm for}\ x>0 \ea\rt.
\eeq
and $v_p$ and $w_p$ defined by
\beq\label{vp}
	v_p \equiv \frc{\tau^2/2 + i(p-\ep)}{-\tau^2/2 + i(p-\ep )},\quad w_p \equiv v_p-1 = \frc{\tau^2}{-\tau^2/2 + i(p-\ep)},
\eeq
maps doubly infinite square-summable sequences $(a_p:p)\equiv (a_p:e^{2ipL} = \b{v}_p)$ onto $L^2(\mathbb{S}) \oplus \C$. It is a simple matter to see that sequences with only one non-zero element $a_p$ yield solutions $g=e_p$, $h=i w_p/\tau$ to (\ref{schro}). Note that the allowed values of $p$ are obtained using the periodicity condition
\begin{eqnarray} \label{percond}
	e^{2ipL}=\bar{v}_p.
\end{eqnarray} 
There is one eigenvalue $p$ in each interval $[\frac{(2k-1)\pi}{2L},\frac{(2k+1)\pi}{2L}]$ except for the two intervals containing $\pm\tau^2/2$ which each contains two eigenvalues. We also show in Appendix \ref{apphilbert} that the completion of the space of such solutions $e_p\oplus iw_p/\tau$, under the usual inner product
\beq\label{inprod}
	\langle g_1 \oplus h_1\;,\; g_2 \oplus h_2\rangle = \int_{-L}^Ldx\, \bar g_1(x)g_2(x) + \bar h_1h_2,
\eeq
is the Hilbert space $L^2(\mathbb{S})\oplus \C$, and corresponds, under the map (\ref{basis}), to the space of doubly infinite square-summable sequences $(a_p:p)$. In particular, although every finite linear combination $g\oplus h$ of basis elements $e_p \oplus i w_p/\tau$ has the property that
\beq\label{jump1p}
	g(0^+)-g(0^-) = -i\tau h,
\eeq
this property does not survive completion: the function $g$ and the number $h$ may be chosen independently, in the sense that for any $L^2(\mathbb{S})$ function $g$ and $h\in \C$, there exists a square-summable sequence $(a_p:p)$ that reproduces them under (\ref{basis}).

From the one-particle problem we can solve for the fermionic operators dynamics. Since the set $e_p \oplus i w_p/\tau$ form a basis of eigenvectors of the one-particle Hamiltonian, this dynamics is diagonalised by introducing canonical operators $a_p$ and $b_p$ such that:
\begin{eqnarray}
	\psi_e(x,t) &=& \frac{\sqrt{2\pi}}{2L} \sum_p  e_p(x)e^{-ipt}\, \hat a_p , \nonumber\\
	\psi_o(x,t) &=&  \frac{\sqrt{2\pi}}{2L}\sum_p  e^{ip(x-t)}\, \hat b_p, \label{eqos}\\
	-i\tau\ d(t) &=&  \frac{\sqrt{2\pi}}{2L} \sum_p  w_p e^{-ipt}\, \hat a_p, \nonumber
\end{eqnarray}
with, again, quantification conditions  $e^{2ipL}=1$ in the odd sector and $e^{2ipL}=\bar{v}_p$ in the even sector. From this, we see that the quantity
\beq\label{phase}
	v_p=:e^{i\phi_p}
\eeq
defined in (\ref{vp}) is the phase shift, and that the transmission coefficient is
\beq\label{transmission}
	T(p)=|w_p/2|=\sin(\phi_p/2) = \frc{\tau^2/2}{\sqrt{\tau^2/4+(p-\ep)^2}}.
\eeq

The inverse of the mode solution is
\begin{eqnarray}\label{inverse}
	\hat a_p &=& \big(1+\frac{|w_p|^2}{2L\tau^2}\big)^{-1}\, \Big[ \int_{-L}^L \frc{dx}{\sqrt{2\pi}}\, \psi_e(x)\, \bar e_p(x) -\frc{i\b{w}_p}{\sqrt{2\pi} \tau}d\Big], \\
	\hat b_p &=& \int_{-L}^L \frc{dx}{\sqrt{2\pi}}\, \psi_o(x) e^{-ipx}.
\end{eqnarray}
The canonical commutation relations $\{\psi_i^{\dag}(x),\psi_j(y)\}=\delta_{i;j}\delta(x-y)$ and $\{d^{\dag},d\}=1$ are equivalent to 
\begin{eqnarray*}
\{\hat b^{\dag}_p,\hat b_{p'}\}=\frac{L}{\pi}\, \delta_{p;p'},\quad
\{\hat a^{\dag}_p,\hat a_{p'}\}=\frac{L}{\pi}\,  \big(1+\frac{|w_p|^2}{2L\tau^2}\big)^{-1}\, \delta_{p;p'}.
\end{eqnarray*}
The other anti-commutators vanish.

The Hilbert space is the Fock space over these anti-commutation relations which makes the energy bounded from below. The vacuum $|\vac\ket$ is defined by
\[
	\h a_{-p}|\vac\ket = \h a^\dag_p|\vac\ket = \h b_{-p}|\vac\ket = \h b_p^\dag |\vac\ket = 0 \qquad \forall\;p<0.
\]
Its physical meaning is of course that the Fermi sea of negative-energy states has been filled. In the Hamiltonian, this infinite negative energy must be appropriately shifted away, so that
\[
	H = \frc{\pi}{L} \sum_p \,p\lt[\lt(1+\frc{|w_p|^2}{2L\tau^2}\rt) :\h a^\dag_p \h a_p:\, + \,:\h b^\dag_p \h b_p:\rt]
\] 
where the normal-ordering is with respect to the vacuum $|\vac\ket$. Note that thanks to (\ref{jump1p}), the operators $\psi_e(x)$ and $d$ are related to each other in any finite-energy state by the discontinuity relation
\beq\label{jump}
	\psi_e(0^+)-\psi_e(0^-) \stackrel{H}= -i\tau d
\eeq
(this also holds in states containing infinitely many eigenstates of the Hamiltonian if the coefficients decrease exponentially with the energy). The discontinuity relation comes from the fact that one set of modes is used to describe both $\psi_e$ and $d$~: this is a phenomenon usually referred to as hybridisation.

We also need to take the large $L$ limit. The density of eigenvalue of the one-particle Hamiltonian is uniform in this limit. Defining $a_p = \lim_{L\to\infty} \h a_p$ and $b_p = \lim_{L\to\infty} \h b_p$, the Hamiltonian is simply $H = \int dp \,p \,(a_p^\dag a_p +b_p^\dag b_p)$, and the fermionic operators decompose as:
\begin{eqnarray} \label{psi_mode}
	\psi_e(x,t) &=& \int \frc{dp}{\sqrt{2\pi}}\, e_p(x)e^{-ipt}\, a_p ,\nonumber\\
	\psi_o(x,t) &=& \int \frc{dp}{\sqrt{2\pi}}\, e^{ip(x-t)}\, b_p,\\
	-i\tau\, d(t) &=& \int \frc{dp}{\sqrt{2\pi}}\, w_p e^{-ipt}\, a_p.\nonumber
\end{eqnarray}
The canonical anti-commutation relations give
\beq\label{canap}
	\{a_p^\dag,a_{p'}\} = \{b_p^\dag,b_{p'}\} = \delta(p-p'),
\eeq
other anti-commutators vanishing.

\subsubsection{Full regularisation via a spreading of the impurity}

An alternative way of dealing with the ambiguities of (\ref{schro}) consists in regularising the Hamiltonian (\ref{H}) so that the impurity $d$ interacts with the itinerant fermions on a neighbourhood of the origin. The regularised Hamiltonian is
\beq\label{Hreg}
	H_\text{reg} = -i \int_{-L}^L dx(\psi_e^\dag \p_x \psi_e + \psi_o^\dag\p_x \psi_o) +
	\tau \int_{-L}^Ldx\,  \varphi^{[a]}(x)\, (\psi^\dag_e(x) d + d^\dag \psi_e(x))
	+ \ep\, d^\dag d,
\eeq
where $ \varphi^{[a]}(x):= a^{-1}\varphi(x/a)$ is a mollifier of the Dirac distribution. The function $\varphi$ is real with compact support on $\R$, normalised such that $\int_{-L}^Ldx\,  \varphi(x)=1$. We consider $a$ small enough so that the support of $\varphi^{[a]}$ lies in the open interval $(-L,L)$. We will denote by  $I_a=[I_a^-,I_a^+]$ the closed interval delimited by the maximum/minimum values $I_a^\pm$ of the support of $\varphi^{[a]}$. We shall be interested in the limit $a\to 0^+$ in which $\varphi^{[a]}(x)\to \delta(x)$. With the boundary conditions $g(-L)=g(L)$ and differentiability on $\mathbb{S}$, the one-particle Hamiltonian is clearly Hermitian under the inner product $\bra\cdot,\cdot\ket$ (\ref{inprod}). Hence, by similar arguments as those of Appendix \ref{apphilbert}, it gives rise, after completion, to a Hermitian operator on the full Hilbert space $L^2(\mathbb{S})\oplus\mathbb{C}$. The eigenvector equations (\ref{schro}) for the even sector, with eigenvalue $p$, are replaced by:
\beqa\label{schro_reg}
	-i\p_x g_p(x) + \tau h_p\, \varphi^{[a]}(x) &=& p\, g_p(x) \n
	\tau\, \bra\varphi^{[a]},g_p\ket &=& (p-\ep)\, h_p,
\eeqa
using $\bra g_1,g_2\ket := \bra g_1\oplus 0,g_2\oplus 0\ket$. Let $\xi_p(x):= e^{-ipx} g_p(x)$ and $\varphi^{[a]}_p(x) := e^{-ipx} \varphi^{[a]}(x)$. Eqs. (\ref{schro_reg}) then become $i\partial_x\xi_p(x)=\tau  h\, \varphi^{[a]}_p(x)$ and $\tau\, \bra \varphi^{[a]}_p,\xi_p\ket = (p-\ep)\, h_p$, and their solutions are:
\[
	g_p(x) = -i\tau\, h_p\, [\Phi_p^{[a]}(x)+c_p]\, e^{ipx}\quad
	\text{with}\quad \Phi_p^{[a]}(x):= \int_{I_a^-}^x dy\, \varphi^{[a]}(y) e^{-ipy},
\]
where the integration constant is
\beq\label{cp}
	c_p := \frc{i(p-\ep)-\tau^2 \bra\varphi^{[a]}_p,\Phi_p^{[a]}\ket}{\tau^2\, \overline{\Phi_p^{[a]}(I^+_a)}}
\eeq
The inner product in $c_p$ can re-written, using integration by part, as
\beq\label{cp2}
	\bra\varphi^{[a]}_p,\Phi_p^{[a]}\ket = \frc12 e^{2i p I_a^+} - ip \int_{I_a} dx\, \Phi_p^{[a]}(x)^2 e^{2ipx}.
\eeq
The periodicity relation $g_p(-L)=g_p(L)$ imposes the quantification condition:
\[
	e^{ipL}\, [ c_p + \Phi_p^{[a]}(I_a^+)]= e^{-ipL}\, c_p.
\]
Note that $|1+\Phi_p^{[a]}(I_a^+)/c_p|=1$ because $2\Re\text{e}\,\bra\varphi^{[a]}_p,\Phi_p^{[a]}\ket=|\Phi_p^{[a]}(I^+_a)|^2$. The quantity $1+\Phi_p^{[a]}(I_a^+)/c_p$ is the phase shift incurred through the spread impurity.

The large-$L$ limit can then easily be taken, whereby the distribution of $p$ values is uniform. With appropriate normalisation, we find, for the full operator solution,
\beq\label{gp}
	g_p(x) = \lt[ 1 + \frc{\Phi_p^{[a]}(x)}{c_p}\rt]\,e^{ipx},\quad h_p = \frc{i}{\tau c_p}
\eeq
and
\begin{eqnarray} \label{psi_mode_reg}
	\psi_e(x,t) &=& \int \frc{dp}{\sqrt{2\pi}}\, g_p(x)e^{-ipt}\, a_p ,\nonumber\\
	\psi_o(x,t) &=& \int \frc{dp}{\sqrt{2\pi}}\, e^{ip(x-t)}\, b_p,\\
	-i\tau\, d(t) &=& \int \frc{dp}{\sqrt{2\pi}}\, c_p^{-1} e^{-ipt}\, a_p.\nonumber
\end{eqnarray}
The canonical anti-commutation relations give again (\ref{canap}). Note that the orthonormality relation
\beq\label{orthoreg}
	\int_{-\infty}^\infty dx\,\b g_p(x) g_{p'}(x) + \b h_p h_{p'} = 2\pi \delta(p-p')
\eeq
follows from the eigenvalue problem and the chosen normalisation, using the fact that $|1+\Phi_p^{[a]}(I_a^+)/c_p|=1$. This in particular gives the modes in terms of the fermion fields:
\beqa\label{mode_psi_reg}
	a_p &=& \int_{-\infty}^\infty \frc{dx}{\sqrt{2\pi}} \b g_p(x) \psi_e(x) - \frc{i}{\tau \b c_p} d\n
	b_p &=& \int_{-\infty}^\infty \frc{dx}{\sqrt{2\pi}} e^{-ipx} \psi_o(x).
\eeqa
The inverse orthonormality relations are simply obtained from (\ref{orthoreg}) and from the fact that $g_p\oplus h_p$ form a complete basis on $L^2(\R)\oplus \C$:
\beq
	\int dp\, g_p(x)\b g_p(x') = 2\pi \delta(x-x'),\quad
	\int dp\, h_p \b h_p = 1,\quad \int dp\,g_p(x) \b h_p = 0.
\eeq
The Hamiltonian takes the same form as before, as well as the Hilbert space, in terms of the new operators $a_p$ and $b_p$.

The discontinuity condition (\ref{jump}) does not completely disappear: it is simply smoothed out. It can be written in the form
\beq\label{jumpreg}
	\psi_e(I_a^+) - \psi_e(I_a^-) \stackrel{H_\text{reg}}=
	\int \frc{dp}{\sqrt{2\pi}}\, \lt(e^{ipI_a^+} - e^{ipI_a^-}\rt)\, a_p e^{-ipt}
	-i\tau \int_{I_a} dy \,\varphi^{[a]}(y) d(y-I_a^+),
\eeq
valid in any finite-energy state. Hence, there is again hybridisation.

The limit $a\to0^+$ can be taken, independently of the limit $L\to\infty$ (both limits commute). In the limit $a\to0^+$, we have $\varphi^{[a]}_p(x)\to \delta(x)$ and $\Phi^{[a]}_p(x)\to \Theta(x)$, the Heaviside function. Further, since $|I_a|\to 0$ and $I_a^+\to 0$, expression (\ref{cp2}) makes it clear that $\bra \varphi^{[a]}_p,\Phi_p^{[a]}\ket\to 1/2$. In this limit of vanishing cut-off, the eigenfunctions $g_p \oplus h_p$, with the normalisation chosen above, simplify to $e_p\oplus iw_p/\tau$, hence we recover the solution obtained from the resolved impurity above. Also, in this limit, the regularised discontinuity condition (\ref{jumpreg}) goes to the discontinuity (\ref{jump}).

In summary, the regularised version (\ref{Hreg}) of the Hamiltonian justifies the prescription consisting in defining $g(0)$ in Eq.~(\ref{schro}) as the symmetric sum $(g(0^+)+g(0^-))/2$. The advantage of the regularised version (\ref{Hreg}) is that it allows us to unambiguously define all averages.

\subsubsection{The case $\tau=0$}

In the case $\tau=0$ (the ``free'' case), the solution to the equations of motion can be obtained similarly, with the principal difference that there is no hybridisation. Hence, for the fermion operators $\psi_e,\psi_o$, we simply have to specialise (\ref{eqos}) or (\ref{psi_mode}) to $\tau=0$, whereby both $\psi_e$ and $\psi_o$ take the same form; for the impurity operator, we simply have $d(t) = e^{i\ep t}d$. That is,
\begin{eqnarray} \label{psi_mode_free}
	\psi_e^\text{free}(x,t) &=& \int \frc{dp}{\sqrt{2\pi}}\, e^{ip(x-t)}\, \t a_p ,\nonumber\\
	\psi_o^\text{free}(x,t) &=& \int \frc{dp}{\sqrt{2\pi}}\, e^{ip(x-t)}\, \t b_p,\\
	d^\text{free}(t) &=& e^{i\ep t}d\nonumber
\end{eqnarray}
with anti-commutation relations (\ref{canap}) for $\t a_p$ and $\t b_p$, as well as $\{d^\dag,d\}=1$, other anti-commutators vanishing. The lack of hybridisation is the fact that the corresponding one-particle solution uses a basis of functions in $L^2(\R)\oplus \C$ where the subspaces $L^2(\R)$ and $\C$ are explicitly separated. Recall that at $t=0$, (\ref{psi_mode}) (or  (\ref{eqos})), (\ref{psi_mode_reg}) and (\ref{psi_mode_free}) are just different parametrisations of the same fundamental operators: $\psi^\text{free}_{e,o}(x,0) = \psi_{e,o}(x,0) = \psi_{e,o}(x) $ and $d^\text{free}(0) = d(0) = d$.

\subsection{The large-time limit and the stationary-state density matrix}\label{ssect_limit}

\subsubsection{Description of the density matrices}

The initial density matrix $\rho_d\otimes \rho_{\rm th}$ of (\ref{deflimdens}) is most naturally constructed using mode operators of the $\tau=0$ solution: the Hamiltonians $H_j$ and the number operators $N_j$ are explicitly diagonalised. More precisely, it takes the form
\beq\label{initdensfree}
	\rho_{\rm th} = \exp\lt[-\int dp\,\lt( \beta \,p\,(\t a^\dag_p \t a_p + \t b^\dag_p \t b_p) 
	+ \frc{Wp-V}2\,(\t a^\dag_p \t b_p + \t b^\dag_p \t a_p) \rt)\rt]/(Z_1Z_2).
\eeq
Although the meaning of this density matrix as an operator on a Hilbert space may be subtle to clarify, we will adopt a simpler viewpoint, understanding density matrices as giving rise to linear functionals on an appropriate space of operators. From this viewpoint, the density matrix is an object -- a measure -- that tells us how to associate averages to operators. We do so by using averages of products of mode operators, which are calculated formally using the cyclic property of the trace and the canonical anti-commutation relations. Hence, we define
\beqa
	\Tr(\rho_{\rm th}\, \t a^\dag_p \t a_q) &=&  \frc12 \lt(\frc1{1+ e^{ (\beta+W/2)p - V/2}}+ \frc1{1+ e^{ (\beta-W/2)p + V/2}}\rt)\n
	\Tr(\rho_{\rm th}\, \t a^\dag_p \t b_q) &=&  \frc12 \lt(\frc1{1+ e^{(\beta+W/2)p - V/2}}- \frc1{1+ e^{(\beta-W/2)p + V/2}}\rt)\n
	\Tr(\rho_{\rm th}\, \t b^\dag_p \t b_q) &=&\frc12 \lt(\frc1{1+ e^{(\beta+W/2)p - V/2}}+ \frc1{1+ e^{(\beta-W/2)p + V/2}}\rt)
	\label{tracemodesfree}
\eeqa
along with application of Wick's theorem for averages of products of more operators.

The non-equilibrium steady-state density matrix $\rho_{\rm stat}$ given in (\ref{rhostat}) is defined similarly; it is expressed most easily using the $\tau\neq0$ mode operators $a_p$ and $b_p$ (both in the resolved regularisation and in the spreading regularisation). Indeed, these operators diagonalise the interacting Hamiltonian $H$, and a solution to the equations (\ref{defYX}) defining the operators $Y$ and $X$ is
\beq
	Y = \frc12 \int dp\,(a^\dag_p b_p + b^\dag_p a_p),\quad X = \frc12 \int dp\,p\,(a^\dag_p b_p + b^\dag_p a_p).
\eeq
Again, using canonical anti-commutation relations and cyclic property, the functional evaluating averages under the density matrix $\rho_{\rm stat}$ can be defined. Of course, by construction we find the same formulae as in (\ref{tracemodesfree}):
\beqa
	\Tr(\rho_{\rm stat} \,a^\dag_p a_q) &=& \Tr(\rho_{\rm th}\, \t a^\dag_p \t a_q)\n
	\Tr(\rho_{\rm stat} \,a^\dag_p b_q) &=& \Tr(\rho_{\rm th}\, \t a^\dag_p \t b_q) \n
	\Tr(\rho_{\rm stat} \,b^\dag_p b_q) &=&\Tr(\rho_{\rm th}\, \t b^\dag_p \t b_q)
	\label{tracemodes}
\eeqa
along with application of Wick's theorem for products of more operators.

\subsubsection{Uniqueness of the (generalised) Hershfield operators}

In order to show uniqueness of the steady-state density matrix, constructed using the operators $X$ and $Y$ as defined around (\ref{defYX}), we only have to show that for any operator $W$ such that $[W,H]=0$ and $[W,\psi_e(x)]=[W,\psi_o(x)]=0$ for all $x$ small enough, we have $W\propto {\bf 1}$. It is sufficient to show that $[W,a_p]=[W,b_p]=0$ for all $p\in\R$, where $a_p$ and $b_p$ are the modes of the solution (\ref{psi_mode}) or (\ref{psi_mode_reg}) for the fermion operators in the resolved or spread impurity regularisation. Thanks to $[W,H]=0$, we have $[W,\psi_e(x,t)]=[W,\psi_o(x,t)]=0$ for $x$ small enough, and for all $t\in\R$. Hence, in both regularisations, we find (with $y=x-t$)
\[
	\int dp \,[W,a_p] e^{ipy} = \int dp \,[W,b_p] e^{ipy}= 0\quad\forall\;y\in\R.
\]
Fourier transforming in $y$ gives the result. This indeed shows that the definition given around (\ref{defYX}) is a good definition. Note that we only need to have one value of $x$, small enough, where (\ref{defYX}) holds. Note also that this proof immediately holds as well in interacting impurity models (with free-fermion baths), because for $x$ small enough, the modes of the fermion operators are the asymptotic state creation operators \cite{Ben?}.

\subsubsection{Existence and description of the large-time, large-$L$ limit}

We now show that the limit (\ref{deflimdens}) exists in a weak sense.
\begin{propo}\label{propolocal}
Define a finitely supported operator $G$ as a finite linear combination of finite products of the form $A\,\prod_{i=1}^k \Or_i(x_i)$, with $\Or_i\in\{\psi_e,\psi_o,\psi_e^\dag,\psi_o^\dag\}$ and $A\in\{{\bf 1},d,d^\dag,d^\dag d\}$. Then the limit (\ref{deflimdens}) exists in a weak sense, and is described by the non-equilibrium steady-state density matrix $\rho_{\rm stat}$.
\beq\label{limitloc}
	\lim_{t_0\to-\infty} \lim_{L\to\infty} \Tr\lt(U_{0;t_0} (\rho_d\otimes \rho_{\rm th})  U_{0;t_0}^\dag G\rt)
	=\Tr \lt(\rho_{\rm stat} G\rt).
\eeq
\end{propo}
We show these statements both in the resolved-impurity regularisation, and in the spread-impurity regularisation. A nontrivial aspect is that the discontinuity equation (\ref{jump}) holds in the resolved-impurity case in the trace above at $t_0<0$ but not at $t_0=0$; from the spread-impurity calculation, we also show that the discontinuity equation arises according to two time scales, one being the spreading length $|I_a|$, where the discontinuity is obtained up to terms of order $|I_a|$, the other being the time scale characterising the approach to the steady state, of the order of $2/\tau^2$ when $|I_a|$ is small. In particular, for the resolved impurity, the discontinuity appears immediately.

\noindent $\bullet$ {\bf Resolved impurity:}

The proof relies on writing the solution to the equations of motion with $\tau\neq 0$ in terms of local operators. It is a simple matter to see that the equations
\beqa
	-i\p_t \psi_o(x,t) &=& i\p_x \psi_o(x,t)\n
	-i \p_t \psi_e(x,t) &=& i\p_x \psi_e(x,t) - \tau \delta(x) d(t) \n
	-i \p_t d(t) &=& -\tau \psi_e(0,t) - \ep d(t),
\eeqa
where $\psi_e(0,t) = (\psi_e(0^+,t)+\psi_e(0^-,t)/2$ for any $t>0$, is solved as follows:
\beqa
	\psi_o(x,t) &=& \psi_o(x-t) \n
	\psi_e(x,t) &=& \psi_e(x-t) -i\tau \Theta(x)\Theta(t-x)\lt(
	-i\tau \int_0^{t-x}dx'\,\psi_e(x'+x-t)e^{-bx'} + e^{-b(t-x)} d\rt) \n
	d(t) &=& -i\tau\int_0^t dx'\,\psi_e(x'-t) e^{-bx'} + e^{-bt}d \label{solloc}
\eeqa
for $0<t<2L$ and where $b:=\tau^2/2+i\ep$. Note that the discontinuity relation (\ref{jump}) holds exactly for all $t>0$. In the expression
\beq \label{Gt0}
	\Tr\lt(U_{0;t_0}(\rho_d\otimes \rho_{\rm th})  U_{0;t_0}^\dag\, G\rt)
	= \Tr\lt((\rho_d\otimes \rho_{\rm th})\,  G(-t_0)\rt)
\eeq
we have all local operators in $G(-t_0)$ evolved at time $t=-t_0$. This is evaluated using the solution (\ref{solloc}), and using the form (\ref{initdensfree}) of the initial thermal density matrix along with the expressions (\ref{psi_mode_free}). Note that the impurity operator part simply factorises, and that the electronic part can be evaluated using formulas (\ref{tracemodesfree}). The result is a linear combination of multiple integrals over positions $x'$ of finite products of local operators $\psi_e(\cdot),\psi_o(\cdot),d$. The integrands of the resulting integrals may have poles where the positions of the local operators collide. The prescription for the evaluation of the integrals is to deform the integration contour in such a way that the complex positions of operators have negative imaginary parts that are in increasing (towards the positive imaginary direction) order from the left to the right. This prescription guarantees the correct ordering of the product of operators and that the local-operator solution is in agreement with the mode solution discussed above.

The limit $L\to\infty$ can easily be taken, and the only effect is to make the explicit local-operator solution (\ref{solloc}) valid for all of $0<t<\infty$. Then, the limit $t_0\to-\infty$ can then also be taken. For any given $G$, there is a (negative) value of $t_0$ large enough such that the factor $\Theta(t-x)$ in the solution (\ref{solloc}) for $\psi_e(x,t)$ is, when the solution is put into $G(-t_0)$, always 1. Also, since correlation functions of local operators in the density matrix $\rho_d\otimes \rho_{\rm th}$ do not grow at large separations (in fact, go to constants), and since the integrand in the solution (\ref{solloc}) has an exponentially decaying factor $e^{-bx}$, the large-time limit of the multiple integrals exist. Since, moreover, the operator $d$ in the solution (\ref{solloc}) always come with an exponentially decaying factor $e^{-bt}$, all terms containing it on the right-hand side of (\ref{solloc}) can be omitted in the large-time limit. Hence, in $\rho_d\otimes \rho_{\rm th}$, we may omit $\rho_d$. As a result, since the ensuing correlation functions only contain the local operators $\psi_e(x-t)$, $\psi_e(x'+x-t)$ and $\psi_e(x'-t)$ (at various positions $x$ and with integrations under $x'$), by space-translation invariance of the density matrix $\rho_d\otimes \rho_{\rm th}$ and of the contour-shift prescription, we may omit the term $-t$ in the arguments of all local operators in (\ref{solloc}). Hence, the limit $L\to\infty$ and $t_0\to-\infty$ of the average of $G(-t_0)$ can be obtained from the modified, infinite-time local-operator solution
\beqa
	\psi_o(x,\infty) &\mapsto& \psi_o(x) \n
	\psi_e(x,\infty) &\mapsto& \psi_e(x) -\tau^2 \Theta(x)
	\int_0^{\infty}dx'\,\psi(x'+x)e^{-bx'}  \n
	d(\infty) &\mapsto& -i\tau\int_0^\infty dx'\,\psi_e(x') e^{-bx'}.
\eeqa
Replacing $\psi_e(\cdot)$, $\psi_o(\cdot)$ by their $\tau=0$ mode expression (\ref{psi_mode_free}), we find
\beqa
	\psi_o(x,\infty) &\mapsto& \int \frc{dp}{\sqrt{2\pi}}\, e^{ipx}\, \t b_p,\\
	\psi_e(x,\infty) &\mapsto& \int \frc{dp}{\sqrt{2\pi}}\, e_p(x)\, \t a_p ,\nonumber\\
	-i\tau\, d(t) &\mapsto& \int \frc{dp}{\sqrt{2\pi}}\, w_p \, \t a_p.\nonumber
\eeqa
We see that the right-hand sides are exactly the expressions (\ref{psi_mode}) for the non-zero-$\tau$ solution. Since these must be evaluated against $\rho_{\rm th}$, and since the evaluation of $\tau=0$ modes against $\rho_{\rm th}$, (\ref{tracemodesfree}), gives the same answer as that of non-zero-$\tau$ modes against the non-equilibrium density matrix $\rho_{\rm stat}$, (\ref{tracemodes}), this proves that the limit $L\to\infty$, $t_0\to-\infty$, in this order, of $\rho_{\rm th}$ gives  $\rho_{\rm stat}$ when evaluated against finitely supported operators.

\newpage

\noindent $\bullet$ {\bf Spread impurity:}

The equations of motion now take the form
\beqa
	-i\p_t \psi_o(x,t) &=& i\p_x \psi_o(x,t) \n
	-i\p_t \psi_e(x,t) &=& i\p_x \psi_e(x,t) -\tau \varphi^{[a]}(x) d(t)\n
	-i\p_t d(t) &=& -\tau \int_{I_a} dx\,\varphi^{[a]}(x) \psi_e(x,t) - \ep d(t).
\eeqa
A solution is obtained through
\beqa\label{sollocreg}
	\psi_o(x,t) &=& \psi_o(x-t) \n
	\psi_e(x,t) &=& \psi_e(x-t) -i\tau \int_{I_a}dy\, \varphi^{[a]}(y) \Theta(x-y)\Theta(t-x+y) d(t-x+y) 
\eeqa
for $0<t<2L$, where $d(t)$ satisfy the integro-differential equation
\[
	-i \p_t d(t) = -\tau \int_{I_a} dx'\,\varphi^{[a]}(x') \psi_e(x'-t) + i\tau^2 \int_0^t dx'\, \phi^{[a]}(x') d(t-x') - \ep d(t)
\]
with
\beq
	\phi^{[a]}(x') = \int_{I_a} dy\, \varphi^{[a]}(x'+y)\varphi^{[a]}(y).
\eeq
This integro-differential equation can be solved recursively as a power series in $\tau$; the power series has an infinite radius of convergence, for any finite $t$ and when evaluated inside averages with other local operators. The structure of this solution is not so important, except for the fact that the limit $L\to\infty$ can, as before, be taken easily, the only effect being that the solution becomes valid for all $t>0$.

Recall that from Eq.~(\ref{Gt0}) we first have to evaluate expectations specified by the initial density matrix $\rho_d\otimes\rho_\text{th}$ of local operators $G(-t_0)$ evolved at time $t=-t_0$ and, then to take the limit $t_0\to-\infty$. So we have to look at the behavior of local operators at large time $t\to+\infty$.

It is of course a simple matter to obtain a more explicit form of the $L=\infty$ solution, using the time-evolved operators $\psi_e(x,t)$ and $d(t)$ in (\ref{psi_mode_reg}) along with the expressions (\ref{mode_psi_reg}) of the mode operators in terms of the initial conditions:
\beqa
	\psi_e(x,t) &=& \frc1{2\pi}\int dx' \lt[ \int dp\, \b g_p(x') g_p(x) e^{-ipt} \rt] \psi_e(x') -\frc{i}{2\pi \tau} \lt[\int dp\,
	\frc{g_p(x)}{\b c_p}  e^{-ipt}\rt] d \n
	d(t) &=&  \frc{i}{2\pi \tau} \int dx' \lt[ \int dp\, \frc{\b g_p(x')}{ c_p} e^{-ipt} \rt] \psi_e(x') + \frc1{2\pi}\lt[\int dp\,
	\frc{1}{\tau^2 |c_p|^2} e^{-ipt}\rt] d.
\eeqa
The explicit solutions (\ref{gp}) and (\ref{cp}) make it clear that as ${\rm Im}(p)\to -\infty$, the expressions $g_p(x)/\b{c}_p e^{-ipt}$ and $1/|c_p|^2 e^{-ipt}$ vanish exponentially if $t>|I_a^+|+|I_a^-|+|x|$. Hence, in the solutions above, the integrals in the part proportional to the operator $d$ can be evaluated by such a contour deformation. The result is, as a function of time $t$, exponentially decaying. Hence, at large time, the part in $d$ vanishes, so that we may restrict ourselves to the part containg $\psi_e(x')$ (and in $\rho_d\otimes \rho_{\rm th}$, we may omit $\rho_d$).

Note that the leading exponential of the large-time decay is of the form $e^{-b_a t}$ where the (positive) real part of $b_a$ is the (absolute value of the) imaginary part of the position of the first zero of $c_p$ away from the real axis. The expression (\ref{cp}) indicates that, for $a$ small enough, this will be near to $\tau^2/2$ (more precisely, $\lim_{a\to0} b_a= \tau^2/2+i\ep$).

In order to correctly evaluate the part containing $\psi_e(x')$, we follow the technique used in the resolved-impurity case, and shift $x'$ to $x'+t$, without changing the result of averages. Hence, we have
\beqa
	\psi_e(x,t) &\mapsto& \frc1{2\pi}\int dx' \lt[ \int dp\, \b g_p(x') g_p(x) e^{-ipt} \rt] \psi_e(x'+t) \n
	d(t) &\mapsto&  \frc{i}{2\pi \tau} \int dx' \lt[ \int dp\, \frc{\b g_p(x')}{ c_p} e^{-ipt} \rt] \psi_e(x'+t) \no
\eeqa
as $t\to\infty$. The limit of large $t$ can now be evaluated by replacing $\psi_e(x)$ by its free-mode expansion,
\beqa
	\psi_e(x,t) &\mapsto& \frc1{(2\pi)^{3/2}}\int dq\, \lt[\int dp \int dx' \, \b g_{p+q}(x') g_{p+q}(x) e^{iqx'-ipt} \rt] \t a_q \n
	d(t) &\mapsto&
	\frc{i}{(2\pi)^{3/2} \tau}\int dq \,\lt[\int dp \int dx'  \, \frc{\b g_{p+q}(x')}{ c_{p+q}} e^{iqx'-ipt} \rt] \t a_q.
	\label{step1}
\eeqa
We evaluate the integrals over $x'$ as follows:
\beqa
	\int dx' \, \b g_{p+q}(x') e^{iqx'} &=& \lt(\int_{I_a} 
	+ \int_{-\infty}^{I_a^-}  + \int_{I_a^+}^{\infty}\rt) dx' \, \b g_{p+q}(x') e^{iqx'} \n
	&=& \int_{I_a} dx' \, \b g_{p+q}(x') e^{iqx'}
	+ \frc{e^{-ipI_a^-}}{-ip+0^+} + \lt(1+\frc{\Phi_{p+q}^{[a]}(I_a^+)}{c_{p+q}}\rt) \frc{e^{-ipI_a^+}}{ip+0^+} \no
\eeqa
By the same argument as above, since $g_p\propto 1/c_p$, the integration over $p$ of the first integral, occuring when it is put back into the expressions in the square brackets in (\ref{step1}), gives a quantity that vanishes exponentially as $t\to\infty$ (again, as $e^{-b_a t}$), hence can be neglected. Further, the last term is subject to the same argument, since the additional explicit pole is at $p=i0^+$, hence is not taken as ${\rm Im}(p)\to-\infty$. The only non-vanishing term as $t\to\infty$ is the one occurring from the pole at $p=-i0^+$ in the second term. The result is
\begin{eqnarray}\label{versinfty}
	\psi_o(x,\infty) &\mapsto& \int \frc{dp}{\sqrt{2\pi}}\, e^{ip(x-t)}\, \t b_p,\nonumber\\
	\psi_e(x,\infty) &\mapsto& \int \frc{dp}{\sqrt{2\pi}}\, g_p(x)e^{-ipt}\, \t a_p ,\\
	-i\tau\, d(\infty) &\mapsto& \int \frc{dp}{\sqrt{2\pi}}\, c_p^{-1} e^{-ipt}\, \t a_p.\nonumber
\end{eqnarray}
By the same argument as in the previous paragraph, this reproduces (\ref{psi_mode_reg}) if the density matrix $\rho_{\rm th}$ is replaced by $\rho_{\rm stat}$, hence shows the proposition.

Finally, we see from the solution (\ref{sollocreg}) that the discontinuity condition (\ref{jumpreg}) is replaced by
\[
	\psi_e(I_a^+,t)-\psi_e(I_a^-,t) 
	= \psi_e(I_a^+-t) - \psi_e(I_a^--t)-i\tau \int_{I_a} dy\, \varphi^{[a]}(y) \Theta(t-I_a^++y) d(t-I_a^++y).
\]
Hence, at $t=0$ we recover $\psi_e(I_a^+) - \psi_e(I_a^-)$ on the right-hand side, whereas at $t>|I_a|$ we recover something that looks similar to (\ref{jumpreg}), except for the first two terms. The discussion above, showing that as $t\to\infty$ we can re-interpret the free modes $\t a_p$ as the modes $a_p$, implies that the first two terms tend to the first integral of (\ref{jumpreg}). Hence, we see two time scales: the time scale $|I_a|$, after which the main part of the spread-impurity discontinuity is reached up to an error of order $|I_a|$, and the time scale $1/{\rm Re}(b_a)\sim 2/\tau^2$ necessary for this error to go to zero.

\section{Transferred charges and the Levitov-Lesovik formula} \label{sectcharges}
We are going to present the proof of the Levitov-Lesovik formula for the p.d.f. of the transferred charges in the case of the resolved impurity. It is clear that an analogous proof can be done in the case of a spread impurity: the logic will be the same but the notations and the computations will be much more cumbersome (without much gain).

\subsection{Definitions and general comments}
As explained in the introduction, we aim at describing the charge $\Delta(t)\equiv Q(t)-Q$ transferred during a large time interval $t$ in the stationary regime. The charge is chosen to be that in the first reservoir, so that
\[
	Q= N_1=\int dx\, \psi^{\dag}_1(x)\psi_1(x)
\]
and, as usual, $Q(t) = U_{t,0}^\dag Q U_{t,0}$.

Let the system be prepared at time $t_0$, and denote the density matrix at time zero by $\rho_0:=U_{0;t_0}(\rho_d\otimes\rho_{\rm th})U_{0;t_0}^{\dag}$. The first definition (\ref{P}) of the charge p.d.f. gives
\[
	P_{t_0}(\lambda,t) = \sum_{q,q_0=-\infty}^{\infty} e^{i\lambda q }\ 
	\Tr\lt(P_{q_0+q} U_t P_{q_0} \rho_0 P_{q_0} U_t^\dag P_{q_0+q}\rt).
\]
The second definition of the measured transferred charge would be similar but with only one projector $P_{q_0}$. See Eq.~(\ref{P2}).
 
The first sum can done using the relation $\sum_q f(q) P_q = f(Q)$. Since $Q$ has an integer spectrum, the second sum can be dealt \cite{Espo} using the formula $\int_0^{2\pi} d\mu\,e^{i\mu (Q-q)} = 2\pi P_q$, so that 
\beq\label{form}
	P_{t_0}(\lambda,t) = \int_0^{2\pi} \frc{d\mu }{2\pi}\ 
	\Tr\lt( \rho_0\,e^{-i(\lambda/2+\mu) Q} e^{i\lambda Q(t)} e^{-i(\lambda/2-\mu) Q} \rt).
\eeq
The second definition for the generating function yields a similar expression for $P_{t_0}(\lambda,t)$, the only difference being that $\mu$ is set to $0$ inside the trace (so that the $\mu$-integration does not matter).

In order to reach the stationary regime, we need to take the limit $L\to\infty$ then $t_0\to-\infty$ of $P_{t_0}(\lambda,t)$. Clearly, neither $Q$ nor $Q(t)$ for any $t$ is finitely supported; hence, their averages do not admit a large $L$ limit. Physically, this is because the reservoir carries an infinite charge at $L\to\infty$. But the operator $\Theta_{\lambda,\mu}(t)$ defined by
\beq
	e^{\Theta_{\lambda,\mu}(t)}:= e^{-i(\lambda/2+\mu) Q} e^{i\lambda Q(t)} e^{-i(\lambda/2-\mu) Q} 
\label{etheta}
\eeq
is finitely supported\footnote{To be more precise, it lies in a completion of the space of finitely-supported operators, in the sense that it involves a converging infinite series of operators uniformally supported on a fixed finite region. However, although every finite sum converges towards the steady state according to Proposition \ref{propolocal}, we will not go into the details of the convergence of this infinite series towards the steady state, as this is beyond the scope of the present paper.}.

Indeed, first, the charge difference $\Delta(t):=Q(t)-Q$ is finitely supported, with spatial extension of size $v_Ft$, because the term in $Q(t)$ diverging for $L$ large is time independent. More precisely, we find
\beq\label{Delta}
	\Delta(t) = \int_0^t dt'\, i[H,Q](t') = \int_0^t dt'\,\cur_1(t').
\eeq
where
\beq\label{J1}
	{\cal J}_1(t) =  \tau\sqrt{2}\, {\rm Im}\, [ \psi_1(0,t)^{\dag}d]
\eeq
and we see that $\Delta(t)$ is supported on the segment $[-v_Ft,0]$. The large-$L$ limit of this operator can be taken from this expression. In terms of modes, the operator $\Delta(t)$ is
\begin{eqnarray}\label{L-Delta}
\Delta(t)\propto\frac{1}{16\pi L^2}\sum_{p\not= k} (\hat a_p^{\dag}\ \hat b_p^{\dag})
\begin{pmatrix} 1-\bar v_pv_k & 1-\bar v_p \\ 1-v_k & 0 \end{pmatrix} 
\begin{pmatrix} \hat a_k\\ \hat b_k \end{pmatrix}\ \delta_t(p-k),
\end{eqnarray}
(the restriction on $p\not= k$ in the sum is automatically ensured by $|v_k|^2=1$) with
\beq\label{deltat}
	\delta_t(p)=\frac{e^{ipt}-1}{2\pi ip}.
\eeq
One can also verify directly from this expression that the large-$L$ limit exists. However, the calculation is more subtle. In particular, the naive regularisation of the large $L$ limit of $\Delta(t)$ which would consists in transforming the sum into an integral and regularising the pole in $1/(p-k)$ coming from $\delta_t(p-k)$ with an $+i0^+$ prescription, $1/(p-k)\mapsto 1/(p-k\pm i0^+)$, is incorrect.

Second, we write $Q(t) = Q + \Delta(t)$ and use the Baker-Campbell-Hausdorff formula to write the product of exponentials as an exponential of multiple commutators,
\[
	\Theta_{\lambda,\mu}(t) = i\lambda \Delta(t) + \mbox{\ commutators between $Q$ and $\Delta(t)$}.
\]
The commutators are all operators supported on the segment $[-v_Ft,0]$ because $\Delta(t)$ is, and because $Q$ is an integration of a local density. Hence, by Proposition \ref{propolocal}, the average of any function of $\Theta_{\lambda,\mu}(t)$ admits a stationary limit, and
\begin{eqnarray} \label{Pstat}
	P_{\rm stat}(\lambda,t):= \lim_{t_0\to-\infty} \lim_{L\to\infty} P_{t_0}(\lambda,t) =\int  \frc{d\mu }{2\pi}\ 
	\Tr\lt(\rho_{stat}\, e^{\Theta_{\lambda,\mu}(t)}  \rt).
\end{eqnarray} 

Finally, the universal behaviour, where the Levitov-Lesovik formula holds, emerges in the large $t$ limit. This is the limit $t\tau^2 \gg v_F$ since $v_F/\tau^2$ is the damping length in the stationary regime. We want to evaluate the large deviation generating function $F(\lambda)$ for the charge statistics, defined by (\ref{F}).

Note that we shall deal only with the first definition (\ref{P}) for the generating function, so that Eq.~(\ref{Pstat}) applies. But we shall later show that $\Tr\lt( \rho_{stat}\, e^{\Theta_{\lambda,\mu}(t)}  \rt) $ is actually $\mu$-independent at large time, so that the large deviation function is identical for the two definitions.

\subsection{Expression for $\Theta_{\lambda,\mu}(t)$ at large $L$ and large $t$}
In order take the limit $L\to\infty$ in the definition (\ref{etheta}), we must perform the multiplications of the exponentials in terms of explicitly finitely supported operators. Since $Q(t)$ is a bilinear expression in the fermion operators preserving fermion numbers, it lies in a representation space of the Lie algebra associated to the group of linear endomorphisms of the one-particle Hilbert space. Hence, $\Theta_{\lambda,\mu}(t)$ also lies in this representation space. Let us denote by $q$, $q(t)$ and $\theta_{\lambda,\mu}(t)$ the one-particle operators associated to $Q$, $Q(t)$ and $\Theta_{\lambda,\mu}(t)$, respectively. Clearly.
\[
	e^{\theta_{\lambda,\mu}(t)} = e^{-i(\lambda/2+\mu) q} e^{i\lambda q(t)} e^{-i(\lambda/2-\mu) q}.
\]
Further, let $\delta_t$, $\delta_t'$ and $\delta_t''$ be the one-particle operators associated to $\Delta(t)=Q(t)-Q$, to $[Q,\Delta(t)]$ and to $[Q,[Q,\Delta(t)]]$, respectively. These are all finitely supported operators.

\begin{propo} The one-particle operator $\theta_{\lambda,\mu}(t)$ corresponding to $\Theta_{\lambda,\mu}(t)$ is given by
\beq \label{expTheta}
e^{\theta_{\lambda,\mu}(t)} = 1+ c_{\delta} \delta_t + c_{\delta^2} \delta^2_t + c_{\delta'} \delta'_t + c_{\delta''} \delta''_t
\eeq
where
\[\ba{c}
	c_\delta = i\sin \lambda,\quad c_{\delta^2} = -2\sin^2\frc{\lambda}2,\quad
	c_{\delta'} = 2\sin \mu \sin \frc{\lambda}2,\z
 	c_{\delta''} = 4i \sin\lt(\frc{\lambda}4+\frc\mu2\rt) \sin\lt(\frc\lambda4-\frc\mu2\rt) \sin\frc{\lambda}2.
 	\ea
\]
\end{propo}

\noindent {\it Proof.} Notice that $q$ is a projector, $q^2=q$, because on the one-particle subspace, $Q$ is the operator counting the number of electrons, hence on the one-particle Hilbert space, there is a basis that diagonalises with eigenvalues 0 and 1. Similarly, $q(t)=q+\delta_t$ is also a projector, $q(t)^2=q(t)$. Hence,
\[ e^{\theta_{\lambda,\mu}(t)} -1 = (\delta + y_1 q\delta_t + y_2 \delta_t q + y_1 y_2 q\delta_t q)y_3 \]
where $y_1 = e^{-i(\lambda/2+\mu)}-1$,\ $ y_2 = e^{-i(\lambda/2-\mu)}-1$ and $ y_3 = e^{i\lambda}-1$. Using $q^2=q$, we have further $\delta''_t=\delta_t -2 q\delta_t q + \delta_t q$. Using also $(q+\delta_t)^2 = q+\delta_t$, we have $q\delta_t+ \delta_t q = \delta_t (1-\delta_t)$, so that
\begin{eqnarray*}
	2\,q\delta_t &=& \delta_t(1-\delta_t) + \delta'_t\\
	2\,\delta_t q &=& \delta_t(1-\delta_t) - \delta'_t \\
	2\, q\delta_tq &=& \delta_t(1-\delta_t) - \delta''_t.
\end{eqnarray*}
Inserting this set of relations in the previous formula for $e^{\theta_{\lambda,\mu}(t)}$, we obtain Eq.~(\ref{expTheta}). \cqfd
\medskip

Expression (\ref{expTheta}) for the operator $e^{\theta_{\lambda,\mu}(t)}$ is explicitly in terms of local operators, hence the limit $L\to\infty$ can be taken. More precisely, the matrix elements of $\delta_t$, $\delta_t'$ and $\delta_t''$ may be calculated using expressions of $\Delta(t)$, $[Q,\Delta(t)]$ and $[Q,[Q,\Delta(t)]]$ as bilinears in the modes $a_p$ and $b_p$ of the large-$L$ theory obtained from expressions (\ref{Delta}) and (\ref{J1}) for $\Delta(t)$. Using (\ref{solloc}), we may define two operators $A(t)$ and $B(t)$ and express them in the large $L$ limit in terms of the fermionic fields:
\beqa
	A(t) := \lim_{L\to\infty} [Q,d(t)] 
	&=& \frc{i\tau}{\sqrt{2}} \int_{0}^t dx\,\psi_1(x-t) e^{-(\tau^2/2+i\ep_d)x} \label{Ndl} \\
	B^\dag(t) := \lim_{L\to\infty} [Q,\psi_1^\dag(0,t)]  
	&=& \psi_1^\dag(-t) + \lim_{L\to\infty} \lt( \frc{i\tau}{2\sqrt{2}} [Q,d(t)]\rt)^\dag. \label{Npsil}
\eeqa
They can also be expressed in terms of the modes $a_p$ and $b_p$:
\begin{eqnarray*}
	A(t) &=& -\frc{i}{2\tau} \int \frc{dp}{\sqrt{2\pi}}  w_{p} \lt( e^{-ipt}
		- e^{-(\tau^2/2+i\ep_d)t}\rt)(a_{p}+b_{p}) \n
	B(t) &=& \frc1{4\sqrt{2}} \int \frc{dp}{\sqrt{2\pi}} \lt( (w_p+4) e^{-ipt} - w_p e^{-(\tau^2/2+i\ep_d)t}\rt)(a_{p}+b_{p})\no
\end{eqnarray*}
From (\ref{Ndl}) and (\ref{Npsil}), we have $[Q,[Q,d(t)]] = -[Q,d(t)]$ and $[Q,[Q,\psi_1^\dag(0,t)]] = [Q,\psi_1^\dag(0,t)]$, so that we find
\beqa
	\lim_{L\to\infty} [Q,\psi_1^\dag(0,t) d(t)] &=&
		B^\dag(t) d(t) + \psi_1^\dag(0,t) A(t) \n
	\lim_{L\to\infty} [Q,[Q,\psi_1^\dag(0,t) d(t)]] &=&
		B^\dag(t) d(t) + 2 B^\dag(t) A(t) - \psi_1^\dag(0,t) A(t)
\eeqa
From this and the equivalent mode decompositions of $d(t)$ and $\psi_1(0,t)$, we obtain the mode decomposition of $\Delta(t)$ and of its commutators with $Q$.

We need these expressions for $t\to\infty$. In order to obtain the leading large-$t$ limit, we can keep in the expressions for $A(t)$ and $B(t)$ only the purely oscillatory exponential, the other terms lead to exponentially decreasing factors. Hence, upon integration over $t$ and after some algebra, we find the one-particle operators:
\[
	\bra p|\lt[Q,-\frc{i\tau}{\sqrt{2}} \int_0^t dt'\,\psi_1^\dag(0,t') d(t')\rt] |p'\ket \simeq
		\frc18 \delta_t(p-p') \mato{cc} 2w_p & \b{w}_p-w_p \\ w_p-\b{w}_p & -2w_p \matf +O(e^{-t\tau^2/2}).
\]
with $\delta_t(p)=\frac{e^{ipt}-1}{ip}$. Similarly, we have
\[
	\bra p|\lt[Q,\lt[Q,-\frc{i\tau}{\sqrt{2}} \int_0^t dt'\,\psi_1^\dag(0,t') d(t')\rt] \rt] |p'\ket \simeq
		\frc18 \delta(p-p') \mato{cc} w_p-\b{w}_p & -2w_p \\ 2w_p & \b{w}_p-w_p \matf +O(e^{-t\tau^2/2}).
\]
Adding the hermitian conjugates and putting everything together, we obtained the desired expressions:
\beqa \label{delta'etal}
	\bra p |\delta_t|p'\ket &=& \frc12 \delta_t(p-p') \mato{cc} 0 & \b{w}_p \\ w_p & 0 \matf +O(e^{-t\tau^2/2})\\
	\bra p |\delta_t^2|p'\ket &=& \frc14 \delta_t(p-p') \mato{cc} -w_p-\b{w}_p & 0 \\ 0 & -w_p-\b{w}_p \matf+O(e^{-t\tau^2/2})\\
	\bra p |\delta_t'|p'\ket &= &\frc14 \delta_t(p-p') \mato{cc} w_p-\b{w}_p & 0 \\ 0 & \b{w}_p  - w_p \matf +O(e^{-t\tau^2/2})\\
	\bra p |\delta_t''|p'\ket&=& \frc14 \delta_t(p-p') \mato{cc} 0 & \b{w}_p  - w_p \\ w_p - \b{w}_p  & 0 \matf +O(e^{-t\tau^2/2}).
	\label{delta"}
\eeqa
We can now put this into (\ref{etheta}) to obtain: 

\begin{propo}\label{propolargeL} In the large $L$, and then large $t$ limit:
\beq \label{thetafin}
	\bra p| e^{\theta_{\lambda,\mu}(t)}-1 |p'\ket \simeq  \delta_t(p-p') \sin\frc\lambda2 \;M(p)+O(e^{-t\tau^2/2})
\eeq
with
\beq\label{mpfin}
	M(p) = e^{i\frac{\mu}{2}\sigma_x}\lt[  i\sigma_y\sin\phi_p-2\sin^2\frac{\phi_p}{2}\,\lt({\bf 1} \sin\frc\lambda2 + i \sigma_x\cos\frc\lambda2 \rt) \rt]e^{-i\frac{\mu}{2}\sigma_x}
\eeq
where $\phi_p$ is defined in (\ref{phase}).
\end{propo}

\noindent {\it Proof.}
Using (\ref{expTheta}) and the previous expressions (\ref{delta'etal}-\ref{delta"}), we write $e^{\theta_{\lambda,\mu}(t)}$ as in Eq.~(\ref{thetafin}) with 
\[
	M(p):= \mato{cc} s_+ c_- w_p + c_+ s_- \b{w}_p & i c_+ c_- \b{w}_p - i s_+ s_- w_p \\
		ic_+ c_- w_p -is_+ s_- \b{w}_p & c_+ s_- w_p + s_+ c_- \b{w}_p \matf
\]
where $s_\pm := \sin\lt(\frc{\lambda}4 \pm \frc\mu2\rt)$ and $c_\pm := \cos\lt(\frc{\lambda}4 \pm \frc\mu2\rt)$.
Some algebra also gives this in the form
\[
	M(p) = i\sin\phi_p\,(\sigma_z  \sin\mu + \sigma_y\cos\mu )-2\sin^2\frc{\phi_p}2\,\lt({\bf 1} \sin\frc\lambda2 +
		i \sigma_x\cos\frc\lambda2 \rt),
\]
which is equivalent to Eq.~(\ref{mpfin}). \cqfd

\subsection{Determinant formula} 
We want to evaluate the large deviation generating function $F(\lambda)$ for the charge statistics:
\beq
	F(\lambda) = \lim_{t\to\infty} t^{-1} \log P_{\rm stat}(\lambda,t).
\eeq
using (\ref{Pstat}). The formula obtained above for the one-particle operator $\theta_{\lambda,\mu}(t)$ correponding $\Theta_{\lambda,\mu}(t)$ can be put to good use from standard free fermion technology. We rewrite $P_{\rm stat}(\lambda,t)$ as a determinant as follows.

\begin{propo} We have
\beq\label{P-det}
{\rm Tr} ( e^{\Theta_{\lambda,\mu}(t)}\rho_{\rm stat})= 
\frac{ {\rm det}(1+ e^{\theta_{\lambda,\mu}(t)} n_{\rm stat} )}{ {\rm det}(1+ n_{\rm stat})}
= {\rm det}( 1 + G_{\rm stat}( e^{\theta_{\lambda,\mu}(t)}-1)) 
\eeq
where $\theta_{\lambda,\mu}(t)$ and $n_{\rm stat}$ are the one-particle operators associated to $\Theta_{\lambda,\mu}(t)$ and $\rho_{\rm stat}$ respectively, and $G_{\rm stat}=\frac{n_{\rm stat}}{1+n_{\rm stat}}$ the matrix of fermion two-point functions.
\end{propo}
\noindent {\it Proof.} Notice that $P_{\rm stat}$ only involves products of operators that are elements of the group of linear endomorphisms of the one-particle Hilbert space. This follows from the discussion above for $e^{\Theta_{\lambda,\mu}(t)}$, and from the fact that the operators $H$, $Y$ and $X$ are exponentiated in $\rho_{\rm stat}$, Eq.~(\ref{rhostat}), are bilinears in fermion operators that preserve the particule number. Further, recall the formula $\Tr(e^M) = \det(1+e^m)$ where $M$ is such a bilinear in fermions, and $m$ is the associated one-particle operator. Here, $\Tr$ is the trace on the full Hilbert space, and ${\rm det}$ is the determinant on the one-particle Hilbert space. Clearly, by the group representation property, this formula extends to product of similar exponentials of bilinear, e.g.~$\Tr(e^Me^N) = \det(1+e^me^n)$ (with obvious notation). Hence, we obtain formula (\ref{P-det}). \cqfd
\medskip

\subsection{Levitov-Lesovik Formula}
Since for $t\to\infty$, the one-particle matrix $\theta_{\lambda,\mu}(t)$ becomes block-diagonal in the momentum basis as in (\ref{thetafin}), with a leading term proportional to $\delta_t(p-p')$, then as $t\to\infty$ we find:
\beq\label{res}
	\log\, {\rm Tr}\ ( e^{\Theta_{\lambda,\mu}(t)}\rho_{\rm stat})= 
		t\, \int \frac{dp}{2\pi}\,\log \lt[\det(1+\h n(p)(e^{\h \theta_{\lambda,\mu}(p)}-1))\rt] + O(1)
\eeq
where the $2\times 2$ matrices $\h \theta_{\lambda,\mu}(p)$ and $\h n(p)$ are the blocks of momentum $p$ of the one-particle matrices associated to the large time asymptotic of $\Theta_{\lambda,\mu}(t)$ and to $G_{\rm stat}$, respectively. The proof that the term $\delta_t(p-p')$ gives rise to the linear dependence at large $t$ upon evaluating the trace (or the determinant) is presented in Appendix \ref{appdelta} (Equation (\ref{resA})).

The matrices involved in (\ref{res}) are, explicitely,
\[
	e^{\h \theta_{\lambda,\mu}(p)}-1 = \sin \frc\lambda2\; M(p),\quad
	\hat n(p) = \frac{1}{2}\big[(n_1(p)+n_2(p)){\bf 1} + (n_1(p)-n_2(p))\sigma_x\big]
\]	
with $n_j(p)$ the thermal occupation numbers of the reservoirs, defined in (\ref{occupation}). Since $\h n(p)$ commutes with $\sigma_x$, the form of $M(p)$ in (\ref{mpfin}) makes it clear that the leading term in (\ref{res}) is independent of $\mu$. Hence, the two different ways (\ref{P}) and (\ref{P2}) of defining the measure of transferred charges give identical large $t$ limit, and the large-deviation function is thus given by:
\beq\label{Flambda}
	F(\lambda) = \int \frac{dp}{2\pi} \log \big[\det(1 + \sin\frac{\lambda}{2}\, M(p) \hat n(p))\big].
\eeq
This shows the following:
\begin{propo}
The large-deviation function $F(\lambda)$, both as defined in (\ref{P}) and in (\ref{P2}), is given by the Levitok-Lesovik formula (\ref{LLformula}).
\end{propo}
\noindent {\it Proof.} This follows from (\ref{Flambda}), taking into account the dispersion relation $\omega=p$ and the transmission coefficient (\ref{transmission}). \cqfd

\section{Conclusion}

We have provided a description of the resonant-level model and its non-equilibrium steady state based on Hershfield's non-equilibrium density matrix, showing that this density matrix is obtained for averages of local (more precisely, finitely-supported) observables, and clarifying aspects related to the point-like impurity. We have then given a precise derivation of the Levitov-Lesovik formula based on local obsevables, making it clear how it enters into the framework of Hershfield's non-equilibrium density matrix. A direct extension of this work is an extension the derivation to a perturbative analysis of the (deformed) Lesovik-Levitov formula for the interacting resonant-level model. This is natural because of the local-operator nature of the derivation, well suited to perturbative analysis. On the other hand, it would be very interesting to study the full-counting statistics in interacting integrable models -- for this, the scattering approach, taken by many authors for deriving the Levitov-Lesovik formula, would probably be more useful. Hence, an important step is to understand the relationship between the correct local-observable description of the steady-state and the scattering approach, in order to have a precise scattering-state derivation. We hope to come back to these questions in future works.

\appendix

\section{Large-$t$ behaviour of the regularised delta-function} \label{appdelta}

We wish to evaluate the right-hand side of the last equation of (\ref{P-det}) in the large-$L$, then large-$t$ limit. According to Proposition \ref{propolargeL}, this has the form
\[
	\det (1 + AB(t))
\]
where the determinant is evaluated on the one-particle Hilbert space. This Hilbert space is a tensor product of a space with a basis enumerated by a continuum momentum variable, and a finite-dimensional (``internal'') space. The operators have the form
\[
	\bra p|A|p'\ket = \delta(p-p')A_p,\quad
	\bra p|B(t)|p'\ket = \delta_t(p-p')B_p + O(e^{-t\tau^2/2})
\]
where $A_p$ and $B_p$ are matrices acting on the internal space.

The determinant on a continuous-basis Hilbert space can be defined via the formula
\[
	\log \det (1 + AB(t)) = \Tr \log (1+AB(t))
	= \sum_{k=1}^\infty \frc{(-1)^{k-1}}k \Tr\lt[(AB(t))^k\rt]
\]
where the trace is defined via an integration over the basis states,
\[
	\Tr\lt[(AB(t))^k\rt] = {\rm tr} \lt(\int_{-\infty}^\infty dp\,\bra p| (A B(t))^k |p\ket\rt)
\]
(with ${\rm tr}$ the trace over the internal space). Hence, we have to evaluate
\[
	\int dp \,\bra p|(AB(t))^k|p\ket
	= \int dp_1\cdots dp_{k}\,\prod_{j=1}^{k} A_{p_j}B_{p_j} \delta_t(p_j-p_{j+1})
\]
(with $p_{k+1}\equiv p_1$ and the product being ordered) in the limit where $t\to\infty$. The most direct way is to use Fourier transforms for the momentum dependence (with coefficients being matrices on the internal space). We will show that, in the sense of distributions over $\alpha_j$s,
\beq\label{toshowA}
	\int dp_1\cdots dp_{k}\,e^{i \sum_{j=1}^k \alpha_jp_j}
	\prod_{j=1}^{k} \delta_t(p_j-p_{j+1})=
	t \int \frc{dp}{2\pi} \,e^{i\lt(\sum_{j=1}^k \alpha_j\rt)p} + O(1)
\eeq
as $t\to\infty$. This implies that
\beq\label{resA}
	\log \det (1 + AB(t)) = t \int \frc{dp}{2\pi}\,\det(1+A_pB_p) + O(1)
\eeq
where the determinant on the right-hand side is on the internal space.

In order to show (\ref{toshowA}), we use (\ref{deltat}), in the form
\[
	\delta_t(p) = \frc1{2\pi} \int_0^t dt'\,e^{ipt'}.
\]
Hence, the left-hand side of (\ref{toshowA}) becomes
\[
	\frc1{(2\pi)^k} \int_{0}^{t} dt_1' \cdots dt_k'\,
	\int dp_1\cdots dp_{k}\,e^{i\sum_{j=1}^k p_j(t_j-t_{j-1}+\alpha_j)}.
\]
Evaluating the integrals over $p_j$s in terms of delta-functions, we find that all $t_j$-integrals are cancelled except for one:
\[
	\int_{A}^{t+B} dt'\,
	\delta\lt(\sum_{j=1}^k \alpha_j\rt)
\]
where $A={\rm max}(\alpha_2,\alpha_2+\alpha_3,\ldots,\sum_{j=2}^k \alpha_j)$ and $B={\rm min}(\alpha_2,\alpha_2+\alpha_3,\ldots,\sum_{j=2}^k \alpha_j)$. The large-$t$ behaviour gives the right-hand side of (\ref{toshowA}).

\section{Current fluctuations}

This Appendix is dedicated to show how the statistics of transferred charges is very different from that of the integrated current. As we shall see the latter is non-universal, depending on a high energy cut-off. Although the mean of the integrated current is finite (because it is equal to the mean of the transferred charge), its p.d.f. is made of three peaks, one centred at zero current, and two others centred at opposite positions diverging with the high energy cut-off. Thus measures of the local current on the impurity are highly fluctuating variables.

We shall only deal with the resolved impurity case.
The integrated current over a time interval $t$ is defined by ${\cal I}_t=\int_0^tds\,I_s$ with $I_s=\tau \Im\text{m}\big[\psi^\dag_o(0,s)d(s)\big]$. We are interested in its p.d.f. so we look for the generating function of its cumulants defined as the expectation of $e^{i\alpha {\cal I}_t}$ in the stationary state:
\[F_\text{current}(\alpha):=\text{Tr}\big(\rho_\text{stat}\, e^{i\lambda {\cal I}_t}\big) = \text{Tr}\big(\rho_\text{stat}\, e^{it\lambda\, {\cal I}_0}\big),\]
where we used the stationary property in the last equation. Since ${\cal I}_0$ is bilinear in the fermion operators, the trace can be evaluated using a formula analogous to Eq.~(\ref{P-det}):
\[ F_\text{current}(\alpha)=\det\big[ 1 + G_\text{stat}(e^{it\alpha\, \widehat\i_0}-1)\big],\]
with $\widehat\i_o$ the one-particle operator associated to ${\cal I}_0$ and $G_\text{stat}$ the matrix of fermion two-point functions.

We skip the details of the computation of this determinant which needs to be regularised. Let $\Lambda$ be a large momentum cut-off: $|p|\leq \Lambda$. The result has the following form:
\begin{eqnarray}
F_\text{current}(\alpha) = A_0 + e^{+it\alpha\sqrt{\Lambda\tau^2/4\pi}}\, A_+  + e^{-it\alpha\sqrt{\Lambda\tau^2/4\pi}}\, A_-,
\end{eqnarray}
with amplitudes $A_0$ and $A_\pm$ given by:
\begin{eqnarray*}
A_+-A_-&=& 2\sqrt{\frac{\pi}{\Lambda}}\, \Im\text{m}\langle\psi_o^\dag d\rangle_\Lambda\\
A_++A_-&=& \frac{\pi}{\Lambda}\big[\langle\psi_o^\dag \psi_o\rangle_\Lambda-
2\langle\psi_o^\dag \psi_o\rangle_\Lambda\, \langle d^\dag d\rangle_\Lambda
+ 2\langle\psi_o^\dag d\rangle_\Lambda\langle d^\dag \psi_o\rangle_\Lambda \big]
+\langle d^\dag d\rangle_\Lambda\\
A_0&=&  1 - A_+-A_-
\end{eqnarray*}
where $\langle\cdots\rangle_\Lambda$ are the regularised stationary expectations. So the peaks are centred at $\pm t\sqrt{\Lambda\tau^2/4\pi}$.

\section{The Hilbert space $L^2(\mathbb{S})\oplus \C$}\label{apphilbert}
Consider the functions $e_p(x)$ defined in (\ref{defep}). These functions have the properties
\[
	e_p(x+L) = e_p(x-L),\quad e_p(-x) = \b{e}_p(x)v_p,
\]
as well as
\beq\label{intep}
	\int_{-L}^L dx\,e_{p'}(x) \b{e}_p(x) = \lt\{\ba{ll} 2L & (p= p') \z -\frc{w_{p'}\b{w}_p}{\tau^2} & (p\neq p'), \ea\rt.
\eeq
where $v_p$ and $w_p$ are define in (\ref{vp}).

Consider the linear space ${\cal S}$ of sequences $(a_p: e^{2ipL}=\bar{v}_p)$ (which we will also denote simply by $(a_p:p)$) satisfying the asymptotic condition $a_p = o(p^{-1})$ as $|p|\to\infty$. Consider also the subspace ${\cal F}$ of $L^2(\mathbb{S})\oplus \C$, with elements which we will denote $g\oplus h$, characterised by the fact that the wave functions $g$ are continuous everywhere except for a finite jump at 0, along with the linear relation $g(0^+)-g(0^-) = -i\tau h$ (here we do not need to specify the value of $g$ at 0). We now show that
\beq\label{defgh}
	g(x) =\sum_p e_p(x) a_p,\quad h = \frc{i}{\tau} \sum_p w_p a_p
\eeq
is a one-to-one map ${\tt M}: {\cal S}\to {\cal F},\;(a_p:p)\mapsto g \oplus h$.

To show that ${\tt M}$ maps ${\cal S}$ into ${\cal F}$, we evaluate, for $\eta>0$ small and $x\neq0$,
\[
	|g(x+\eta)-g(x-\eta)| \leq 2 \sum_p |a_p| \,|\sin p\eta| \leq
	2\eta^{1/2} \sum_{|p|<\eta^{-1/2}} |a_p| + 2\sum_{|p|\ge\eta^{-1/2}} |a_p|.
\]
The right-hand side tends to 0 as $\eta\to0$, since the sum of $|a_p|$ is convergent. Further, the case $x=0$ gives
\[
	g(\eta)-g(-\eta) = \sum_p (v_p e^{ip\eta} - e^{-ip\eta}) a_p
		= \sum_p w_p a_p e^{ip\eta} + 2i \sum_p a_p\, \sin p\eta.
\]
By the same arguments as before, the last sum on the right-hand side of the last equation tends to 0, and by convergence of the sum defining $h$ in (\ref{defgh}), the first sum tends to $-i \tau h$. To show that the map is one-to-one, we simply inverse the equations (\ref{defgh}). By interchanging sum and integral (which can be done by uniform convergence) and using (\ref{intep}), we see that
\[
	\frc1{2L} \int_{-L}^L dx\,g(x) \b{e}_p(x) = a_p +
		\frc1{2L} \frc{\b{w}_p(i \tau h + w_pa_p))}{\tau^2}
\]
hence
\beq\label{defap}
	a_p = \frc1{2L+\frc{|w_p|^2}{\tau^2}} \lt(
	\int_{-L}^L dx\,g(x) \b{e}_p(x) - \frc{i h\b{w}_p}\tau\rt).
\eeq

Consider the inner product on ${\cal F}$ induced by the standard one on $L^2(\R)\oplus \C$:
\beq\label{normgh}
	(g\oplus h,g'\oplus h') = \int_{-L}^L dx\,g(x) \b{g}'(x) + h \b{h}'.
\eeq
This in turn induces an inner product on ${\cal S}$ (which we will likewise denote by $(\cdot,\cdot)$), obtained by the mapping (\ref{defgh}) via $((a_p:p),(a_p':p)) = ({\tt M}(a_p:p),{\tt M}(a_p':p))$. Using (\ref{intep}), we find
\beq
	((a_p:p),(a_p':p)) = \sum_{p} \lt(2L +\frc{|w_p|^2}{\tau^2}\rt) a_p \b{a}_p'.
\eeq
We see that (\ref{defap}) can be written as
\beq
	a_p = \frc1{2L+\frc{|w_p|^2}{\tau^2}} (g\oplus h,e_p\oplus i w_p/\tau).
\eeq
Likewise, by inserting this into (\ref{defgh}), we find that the vectors $\delta_x$ and $\omega$ (the former having a wave-function part that is in fact a distribution) defined by
\beq
	\delta_x := \sum_p \frc{\b{e}_p(x) (e_p \oplus i w_p/\tau)}{2L + \frc{|w_p|2}{\tau^2}},\quad
	\omega := \sum_p \frc{\b{w}_p (e_p \oplus w_p/\tau)}{2L + \frc{|w_p|2}{\tau^2}}
\eeq
have the properties
\beq
	g(x) = (g\oplus h,\delta_x),\quad h = (g\oplus h,\omega).
\eeq

Since the map ${\tt M}$ is one-to-one, we see, by the definitions of the inner products, that it is an isomorphism of the inner product spaces ${\cal S}$ and ${\tt M}({\cal S})\subset {\cal F}$. The completions of two isomorphic inner product spaces are isomorphic Hilbert spaces, and the map ${\tt M}$ extends to an isomorphism of these completions. The completion of ${\cal S}$ is the Hilbert space of square-summable sequences. We now show that the completion of ${\tt M}({\cal S})$ is the Hilbert space $L^2(\mathbb{S})\oplus \C$. The main argument is to show that this completion must contain a direct sum of functions with an arbitrary discontinuity at 0, plus $\C$; that is, that the linear relation between $g$ and $h$ disappears. The completion of this direct-sum space under the direct-sum norm (\ref{normgh}) is $L^2(\mathbb{S})\oplus \C$. To show the former assertion, we may form Cauchy sequences of functions in ${\cal F}$ such that the limit is a function where the discontinuity at 0 is changed from $h$ to another value -- such sequences are straightforward to find. However, since we haven't shown that ${\tt M}({\cal S}) = {\cal F}$, it is not evident that the sequence will lie in ${\tt M}({\cal S})$ without explicitly exhibiting an example. Consider the mapping (\ref{defgh}), with sequences $(a_p:p)$ that are of the form $a_p = c/p + o(p)$ as $p\to\pm\infty$, where $c$ is some complex numbers. These sequences are in the completion of ${\cal S}$, and we may show that they give rise to functions $g$ whose discontinuity at 0 can be any complex number. Indeed, in this case, the discontinuity at 0 can be calculated as follows:
\beqa
	g(\eta)-g(-\eta) &=& \sum_p (v_p e^{ip\eta} - e^{-ip\eta}) a_p
		\sim -i\tau h + \sum_p (e^{ip\eta} - e^{-ip\eta}) \frc{c_0}{p} \n
		&\sim& -i\tau h + 2 i c_0 \int dp\, \frc{\sin p}{p} = -i\tau h + 2\pi i c_0
\eeqa
as $\eta\to0^+$. Likewise, a similar calculation shows that we may put any discontinuities at any finite number of points $x$. The completion of a space of functions continuous everywhere except for an arbitrary number of arbitrary discontinuities is $L^2(\mathbb{S})$. This shows that the completion of ${\tt M}({\cal S})$ is $L^2(\mathbb{S})\oplus \C$.

Finally, it is simple to see that every one-term sequence maps under (\ref{defgh}) to a solution to the equations of motion (\ref{schro}) with finite energy $E=p$. Hence, the completion of the space of finite-energy solutions is indeed the Hilbert space $L^2(\mathbb{S}) \oplus \C$ (with the standard norm).

\end{document}